\documentclass[12pt,dvipdfmx]{article}

\usepackage{amsmath, mathrsfs, bm, amssymb, color, physics, bbm, amsthm}
\usepackage{latexsym}
\usepackage{graphicx}
\usepackage{subcaption} %
\newtheorem{theorem}{Theorem}[section]
\newtheorem{proposition}[theorem]{Proposition}
\newtheorem{lemma}[theorem]{Lemma}
\newtheorem{corollary}[theorem]{Corollary}
\newtheorem{definition}[theorem]{Definition}

\newtheorem{remark}[theorem]{Remark}

\newcommand{\eps}{\varepsilon}

\newcommand{\RR}{\mathbb{R}}
\newcommand{\NN}{\mathbb{N}}
\newcommand{\ZZ}{\mathbb{Z}}
\newcommand{\CC}{\mathbb{C}}
\newcommand{\one}{\mathbbm{1}}

\newcommand{\half}{\frac12}
\newcommand{\cH}{\mathcal{H}}
\newcommand{\Ran}{\mathrm{Ran}}
\newcommand{\Ker}{\mathrm{Ker}}

\newcommand{\spec}{\mathrm{spec}}
\newtheorem*{theorem*}{Theorem}
\makeatletter
\@addtoreset{equation}{section}
\makeatother

\begin{document}
\title{\sc Strong-coupling asymptotics for \\ the anisotropic Rabi model}
\author{
    Masao Hirokawa\footnote{Faculty of Information Science and Electrical Engineering, Kyushu University},  
    Fumio Hiroshima\footnote{Faculty of Mathematics, Kyushu University} and 
    Dong-Yun Lee\footnote{Joint Graduate School of Mathematics for Innovation, Kyushu University}
}
\date{\today}
\maketitle
\begin{abstract}
The anisotropic quantum Rabi model provides a continuous interpolation between the quantum Rabi model and the Jaynes-Cummings model. In this paper, we study the norm-resolvent limit of the anisotropic quantum Rabi model as the coupling strength tends to infinity. After subtracting an explicit quadratic renormalization term, we identify the limiting Hamiltonian, which determines the asymptotic spectral structure of the model. Our result provides a unified description connecting the Jaynes-Cummings model and the full quantum Rabi model in the strong-coupling limit, thereby revealing the dominant role of the counter-rotating interaction.
\end{abstract}
\tableofcontents
\section{Introduction}
The quantum Rabi model~\cite{Rabi36} describes a linear interaction between a two-level quantum system and a single-mode bosonic field and serves as a typical  model in quantum optics.  The quantum Rabi model possesses  $\ZZ_2$-symmetry. 
A standard approximation technique is the so-called rotating-wave approximation (RWA), which yields the Jaynes-Cummings model~\cite{JC63, SK93} by neglecting the counter-rotating terms $\sigma_+ \otimes a^\dagger + \sigma_- \otimes a$ from the quantum Rabi model. Under the RWA, 
the total excitation-number is conserved, and the Hamiltonian possesses the continuous $U(1)$-symmetry generated by the total excitation-number operator.  See \eqref{eq:JC_U1_symm}. In contrast, the full quantum Rabi Hamiltonian retains the counter-rotating terms, breaking the total excitation-number conservation and the $U(1)$-symmetry. %See \eqref{eq:QR_Z2_symm}. 
This structural distinction leads to fundamental differences in spectral properties of  
the quantum Rabi model and the Jaynes-Cummings model. 

In physics, RWA is known to become inadequate in the sufficiently strong coupling regime~\cite{KMLS19, FLRKS19, LeBoite20, CT24}. This provides a motivation for studying the quantum Rabi model, in which the counter-rotating terms are retained.
More generally, the anisotropic Rabi model~\cite{XCCAF14, SYLCW14, TEPG14, ZZ15, Zhang16, ZC17, WXSSX19} assigns independent two coupling parameters $\lambda_1$ and $\lambda_2$ 
to the rotating and counter-rotating interaction terms. This gives a two-parameter family of Hamiltonians $H_{\lambda_1,\lambda_2}$ containing both the Jaynes-Cummings and quantum Rabi Hamiltonians as special cases. 
More precisely $H_{\lambda,\lambda}$ is the quantum Rabi Hamiltonian, whereas 
$H_{0,\lambda}$ is the  Jaynes-Cummings Hamiltonian. 
The precise form of the Hamiltonian considered in this paper is given in Section~\ref{sec:anisotropic}.

From the viewpoint of solving the eigenvalue problem, the Jaynes-Cummings model is exactly solvable due to $U(1)$-symmetry, which allows a reduction of the eigenvalue problem to finite-dimensional subspaces. Although the quantum Rabi model lacks such a reduction, its spectrum can be characterized via the zeros of transcendental functions; see, e.g.,~\cite{Braak11, MPS16}. Similar transcendental characterizations have been extended to the anisotropic Rabi model; see, e.g.,~\cite{XCCAF14, TEPG14}. 
These studies mainly concern spectral properties at any  coupling strengths.

The purpose of the present paper is instead to investigate the strong-coupling limit
$|g|\to~\infty$. 
Strong-coupling asymptotics have been rigorously studied for both the quantum Rabi model~\cite{Hirokawa15, HMS17, RW24, HS25a} and the two-photon quantum Rabi model~\cite{HH26}. See also \cite{bra23,nak25}. To our knowledge, however, no corresponding mathematical analysis has been established for the anisotropic case.

To describe the mathematical structure of this problem, we write the anisotropic Rabi Hamiltonian in terms of the position operator $q$ and the momentum operator $p$ as
\begin{align*}
    K_{\alpha,\beta}(g)
    =
    \begin{pmatrix}
        \triangle+\frac12 (p^2 + q^2) & 0\\
        0 & - \triangle+\frac12 (p^2 + q^2)
    \end{pmatrix}
    +
    g
    \begin{pmatrix}
        0 & \alpha q + i\beta p\\
        \alpha q - i\beta p & 0
    \end{pmatrix}. 
\end{align*}
Here $\alpha,\beta\in\RR$ are parameters and $\triangle\in\RR$ is a nonzero constant.
$\lambda_1$ and $\lambda_2$ are related to $\alpha$ and $\beta$ by 
$\alpha=\frac{\lambda_1+\lambda_2}{\sqrt2}$ and 
$\beta=\frac{\lambda_1-\lambda_2}{\sqrt2}$. 
In this parametrization, the cases $\alpha=0$ or $\beta=0$ correspond to the isotropic Rabi model, while $|\alpha|=|\beta|$ gives the Jaynes-Cummings or anti-Jaynes-Cummings model, whose spectra are explicitly known. We therefore focus on the genuinely anisotropic and non-solvable case
\begin{align*}
    \alpha\beta \neq 0,
    \quad
    |\alpha| \neq |\beta|.
\end{align*}
Moreover, the Hamiltonians $K_{\alpha,\beta}(g)$ and $K_{\beta,\alpha}(g)$ are unitarily equivalent, with the roles of position and momentum interchanged under the Fourier transform. It is therefore enough to analyze the regime $\{(\alpha,\beta)\in\RR^2\mid |\alpha|>|\beta|\}$.

The first step in the strong-coupling analysis is to remove the leading quadratic self-energy and perform a displacement adapted to the dominant $q$-interaction. For $|\alpha|>|\beta|$, after adding $g^2\alpha^2/2$ and applying a suitable unitary transformation 
\begin{align*}\hat{K}_{\alpha,\beta}(g)=
\mathcal{U} _{g\alpha} K_{\alpha,\beta}\mathcal{U} _{g\alpha}^{-1},\end{align*} 
the renormalized Hamiltonian takes the form
\begin{align*}
    \hat{K}_{\alpha,\beta}^{\mathrm{ren} }(g)
    =\hat{K}_{\alpha,\beta}(g)+\frac{g^2\alpha^2}{2}=    \begin{pmatrix}
        \frac12 (p^2 + q^2) & e^{-2ig\alpha p} (-\triangle + ig\beta p)\\
        e^{2ig\alpha p}(-\triangle - ig\beta p) & \frac12 (p^2 + q^2)
    \end{pmatrix}.
\end{align*}
This representation reveals the main difficulty caused by the anisotropy. When $\beta=0$, the off-diagonal interaction is a bounded oscillatory operator. Its weak convergence to zero, together with the compactness of the resolvent of the harmonic oscillator, yields the norm resolvent convergence to the decoupled harmonic oscillator~\cite{Hirokawa15, HMS17,HS25a,RW24}. For $\beta\ne0$, however, the oscillatory factors are accompanied by
\begin{align*}
    ig\beta p e^{\mp2ig\alpha p},
\end{align*}
which involve both the unbounded momentum operator $p$ and the  coefficient $ig\beta$ growing linearly in $g$. Thus, the anisotropic interaction cannot be regarded as a bounded oscillatory perturbation, and the compactness argument used in the isotropic case no longer applies directly.

To overcome this difficulty, we exploit the $2\times2$ operator structure and use Schur complements to represent the resolvent $(\hat{K}_{\alpha,\beta}^{\mathrm{ren} }(g)-z)^{-1}$ in a matrix form. We then analyze the strong-coupling behavior of its diagonal and off-diagonal entries separately. A precise asymptotic analysis shows that the diagonal entries strongly converge  to the resolvent of the {\it effective harmonic oscillator}:
\begin{align}
\label{effho}
\hat h_\infty=    \frac12
    \left(
        q^2 + \left(1 - \frac{\beta^2}{\alpha^2}\right) p^2
    \right),
\end{align}
whereas the off-diagonal entries vanish. The definition of effective harmonic oscillators are given by Definition \ref{eff}. 
In the regime $|\alpha|>|\beta|$, the compactness argument of Lemma~\ref{lem:compact_domination} further upgrades this convergence to norm resolvent convergence. By contrast, in the critical case $|\alpha|=|\beta|$, the limiting resolvent is no longer compact, so the convergence cannot hold in norm and remains at the strong resolvent level. 
Consequently we have the main theorem (Theorem \ref{thm:resolvent_conv}) in this paper. 
\begin{theorem*}
Assume $(\alpha,\beta)\ne(0,0)$.
\begin{enumerate}
\item[(1)] Suppose that $|\alpha|>|\beta|$.
Then 
\begin{align*}
         \hat{K}_{\alpha,\beta}(g) 
        +
        \frac{g^2 \alpha^2}{2}
    \longrightarrow 
    \begin{pmatrix}
        \frac12 \left(q^2 + \left(1 - \frac{\beta^2}{\alpha^2}\right) p^2\right) & 0\\
        0 & \frac12 \left(q^2 + \left(1 - \frac{\beta^2}{\alpha^2}\right) p^2\right)
    \end{pmatrix}
\end{align*}
as $|g|\to\infty$ in the norm resolvent sense.

\item[(2)] Suppose that $|\alpha|=|\beta|$. Then
\begin{align*}
         \hat{K}_{\alpha,\beta}(g) 
        +
        \frac{g^2 \alpha^2}{2}
    \longrightarrow 
    \begin{pmatrix}
        \frac12 q^2 & 0\\
        0 & \frac12 q^2
    \end{pmatrix}
\end{align*}
as $|g|\to\infty$ in the strong resolvent sense.

\item[(3)] Suppose that $|\alpha|<|\beta|$. 
Then 
\begin{align*}
         \check{K}_{\alpha,\beta}(g) 
        +
        \frac{g^2 \beta^2}{2}
\longrightarrow 
    \begin{pmatrix}
        \frac12 \left(p^2 + \left(1 - \frac{\alpha^2}{\beta^2}\right) q^2\right) & 0\\
        0 & \frac12 \left(p^2 + \left(1 - \frac{\alpha^2}{\beta^2}\right) q^2\right)
    \end{pmatrix}
\end{align*}
as $|g|\to\infty$ in the norm resolvent sense.
\end{enumerate}
\end{theorem*}
Here 
\begin{align*}\check{K}_{\alpha,\beta}(g)=
\mathcal{V} _{g\beta} K_{\alpha,\beta}\mathcal{V} _{g\beta}^{-1}
\end{align*} 
with some unitary operator $\mathcal{V} _{g\beta}$.  
See Theorem \ref{thm:resolvent_conv}. 
Thus, in the genuinely anisotropic regime, the interaction does not simply disappear in the strong-coupling limit. Rather, its contribution survives through a reduction of the kinetic term of the effective harmonic oscillator. The resulting norm resolvent convergence, in turn, determines the asymptotic behavior of the eigenvalues of the anisotropic Rabi Hamiltonian. See Corollary \ref{col:eigenvalues_conv}.

Finally, we briefly review the physical background. Although the strict 
strong-coupling limit is a mathematical idealization, recent experimental 
advances have brought the anisotropic quantum Rabi model into the 
deep-strong-coupling regime. In particular, a programmable 
superconducting-circuit realization with independently tunable rotating 
and counter-rotating couplings has revealed an anisotropy-induced 
reconstruction of the level spacings and the resulting 
collapse-and-revival dynamics~\cite{Song26}. These developments provide 
experimental motivation for investigating the asymptotic spectral 
structure in the strong-coupling regime.
The eigenvalues of the effective harmonic oscillator~\eqref{effho} are 
equally spaced, with spacing
\begin{align*}
    \sqrt{1-\frac{\beta^2}{\alpha^2}}.
\end{align*}
Remarkably, the same anisotropy-dependent factor appears in the 
hidden-symmetry energy scale observed in the recent 
experiment~\cite{Song26}.
%Our investigation is primarily motivated by a mathematical  analysis of the strong-coupling limit. 
At the same time, the resulting 
asymptotic spectral structure may also be relevant to the viewpoint that 
strong-coupling phenomena in quantum Rabi-type models exhibit features 
reminiscent of spontaneous supersymmetry breaking; see, 
e.g.,~\cite{bcbs16,Hirokawa15,Hirokawa24,Hirokawa25,TPG15,
CWMZJYHZD22,KMMR24}.

This paper is organized as follows. In Section~\ref{s2}, we define the quantum Rabi model,
the Jaynes-Cummings model, and the anisotropic Rabi model.
Section~\ref{s3} is devoted to the norm resolvent convergence of the anisotropic Rabi Hamiltonian. 
as $|g|\to\infty$. 
In Section \ref{s4} we investigate the asymptotic behavior of the eigenvalues of the anisotropic Rabi Hamiltonian.
\section{Quantum models}\label{s2}
\subsection{$\ZZ_2$-symmetry of quantum Rabi model}
Let $p$ and $q$ be the self-adjoint operators on $L^2(\RR)$, defined by $(pf)(x) = -i \frac{df}{dx}(x)$ and $(qf)(x) = xf(x)$ with domains
\begin{align*}
    D(p) &= \{f \in L^2(\RR) \mid |k|\hat{f}(k) \in L^2(\RR) \}, 
    \\
    D(q) &= \{f \in L^2(\RR) \mid |x|f(x) \in L^2(\RR) \},
\end{align*}
where $\hat{f}$ denotes the Fourier transform of $f$ in $L^2(\RR)$. They satisfy the canonical commutation relation $[q,p] = i$ on $D(pq) \cap D(qp)$. 
We set the harmonic oscillator as 
\begin{align}\label{h}
h=\half (p^2+q^2).
\end{align}
The annihilation and creation operators are defined by 
\begin{align*}
    a = \frac{1}{\sqrt{2}} (q + ip),
    \quad 
    a^\dagger = \frac{1}{\sqrt{2}} (q - ip),
\end{align*}
satisfying 
$    [a, a^\dagger] = 1$ 
on $D(a^\dagger a) \cap D(a a^\dagger)$ and the adjoint of $a$ is given by $a^\ast=a^\dagger$. 
    We have
\begin{align*}
    a^\dagger a + \frac12 = h. 
\end{align*}
We shall frequently use the following unitary equivalence.
For $A,B>0$, we have
\begin{align*}
    \frac12 (Ap^2 + Bq^2)
    \cong
    \sqrt{AB} h.
\end{align*}
In particular,
\begin{align*}
    \operatorname{spec}\left(\frac12 (Ap^2 + Bq^2)\right)
    =
    \left\{
        \sqrt{AB} \left(n + \frac12\right)
    \right\}_{n=0}^\infty.
\end{align*}
The two-level spin degrees of freedom are represented by the $2 \times 2$ Pauli matrices:
\begin{align*}
    \sigma_x = \begin{pmatrix} 0 & 1\\ 1 & 0 \end{pmatrix}, \quad
    \sigma_y = \begin{pmatrix} 0 & -i\\ i & 0 \end{pmatrix}, \quad
    \sigma_z = \begin{pmatrix} 1 & 0\\ 0 & -1 \end{pmatrix}.
\end{align*}
For $g \in \RR$ and $\triangle \in \RR \setminus \{0\}$, the quantum Rabi Hamiltonian acts on the total Hilbert space 
\begin{align*}
    \cH  = \CC^2 \otimes L^2(\RR)
\end{align*} 
and is given by
\begin{align}\label{def:QR}
    H_\mathrm{QR}(g) 
    = 
    \triangle\sigma_z \otimes \one 
    + 
    \one \otimes \left(a^\dagger a + \frac12\right) 
    + 
    g\sigma_x \otimes (a + a^\dagger).
\end{align}
Here   
$|\triangle|$ denotes half the energy-level separation of the two-level system. Throughout this paper, we fix $\triangle$. Let
\begin{align*}
    \Pi = (-1)^{a^\dagger a}
\end{align*}
be the bosonic parity operator, and define the parity operator on $\cH$ by
\begin{align*}
    P = \sigma_z \otimes \Pi.
\end{align*}
Then $P$ is a unitary self-adjoint involution, i.e., $P^\ast=P=P^{-1}$, and the quantum Rabi Hamiltonian satisfies
\begin{align}\label{eq:QR_Z2_symm}
    [H_{\mathrm{QR}}(g), P] = 0,
\end{align}
reflecting its discrete $\ZZ_2$-symmetry.

The quantum Rabi model is one of the simplest nontrivial models describing the interaction between a two-level system and a bosonic mode as is seen in \eqref{def:QR}. 
Despite its apparently simple form, its spectral structure is highly nontrivial and provides a rich source of problems in spectral and operator theory. 
\subsection{$U(1)$-symmetry of JC and AJC models}
Introducing the ladder operators:  
\begin{align*}
    \sigma_+ = \begin{pmatrix} 0 & 1\\ 0 & 0 \end{pmatrix}, 
    \quad
    \sigma_- = \begin{pmatrix} 0 & 0\\ 1 & 0 \end{pmatrix},
\end{align*}
the interaction term expands as
\begin{align*}
    g\sigma_x \otimes (a+a^\dagger)
    = 
    g\left(
        \sigma_+ \otimes a + \sigma_- \otimes a^\dagger 
        + \sigma_- \otimes a + \sigma_+ \otimes a^\dagger
    \right).
\end{align*}
Let 
$
    N_T = \sigma_+\sigma_- \otimes \one + \one \otimes a^\dagger a
$
be the total excitation-number operator. The terms $\sigma_+ \otimes a$ and $\sigma_- \otimes a^\dagger$ commute with $N_T$ and are referred to as the rotating terms, whereas $\sigma_- \otimes a$ and $\sigma_+ \otimes a^\dagger$ do not and are referred to as the counter-rotating terms. Neglecting the counter-rotating terms yields the Jaynes-Cummings (JC) Hamiltonian:
\begin{align*}
    H_\mathrm{JC} (g) 
    = 
    \triangle\sigma_z \otimes \one 
    + 
    \one \otimes \left( a^\dagger a + \frac12 \right) 
    + 
    g(\sigma_- \otimes a^\dagger + \sigma_+ \otimes a). 
\end{align*}
We highlight an additional symmetry of the JC Hamiltonian. Like the quantum Rabi Hamiltonian, the JC Hamiltonian possesses the $\ZZ_2$-symmetry generated by the parity operator $P$. In addition, it satisfies
\begin{align}\label{eq:JC_U1_symm}
    e^{i\theta N_T} H_{\mathrm{JC} }(g) e^{-i\theta N_T}
    = 
    H_{\mathrm{JC} }(g),
    \quad
    \theta \in \RR,
\end{align}
and hence
\begin{align*}
    [H_{\mathrm{JC} }(g), N_T] = 0.
\end{align*}
This additional continuous $U(1)$-symmetry decomposes $\CC^2 \otimes L^2(\RR)$ into finite-dimensional eigenspaces of $N_T$ explicitly yielding the spectrum:
\begin{align*}
    \mathrm{spec}(H_{\mathrm{JC} }(g)) 
    = 
    \left\{ \frac12 - \triangle \right\} 
    \cup 
    \left\{ 
        n \pm \sqrt{ g^2 n + \left( \frac12 - \triangle \right)^2 }
    \right\}_{n=1}^{\infty}.
\end{align*}

Similarly, the anti-Jaynes-Cummings (AJC) Hamiltonian~\cite{SK93, KMMR24, BCNNG24} defined by
\begin{align*}
    H_{\mathrm{AJC}}(g) 
    = 
    \triangle\sigma_z \otimes \one 
    + 
    \one \otimes \left( a^\dagger a + \frac12 \right) 
    + 
    g (\sigma_+ \otimes a^\dagger + \sigma_- \otimes a).
\end{align*}
It is unitarily equivalent to %$H_{\mathrm{JC} }(g)$ via 
\begin{align}\label{eq:JC/AJC_unitary}
    (\sigma_x \otimes \one) H_{\mathrm{AJC}}(g) (\sigma_x \otimes \one) 
    = 
    -\triangle\sigma_z \otimes \one 
    + 
    \one \otimes \left( a^\dagger a + \frac12 \right) 
    + 
    g(\sigma_- \otimes a^\dagger + \sigma_+ \otimes a),
\end{align}
which is the JC Hamiltonian with $\triangle$ replaced by $-\triangle$. Then we have
\begin{align*}
    \mathrm{spec}(H_{\mathrm{AJC}}(g)) 
    = 
    \left\{ \frac12 + \triangle \right\} 
    \cup 
    \left\{ 
        n \pm \sqrt{ g^2 n + \left( \frac12 + \triangle \right)^2 }
    \right\}_{n=1}^{\infty}. 
\end{align*}
Correspondingly, the AJC Hamiltonian also possesses a $U(1)$-symmetry. Let
\begin{align*}
    N_A
    =
    \sigma_-\sigma_+ \otimes \one
    +
    \one \otimes a^\dagger a.
\end{align*}
Then
\begin{align}\label{eq:AJC_U1_symm}
    e^{i\theta N_A} H_{\mathrm{AJC}}(g) e^{-i\theta N_A}
    =
    H_{\mathrm{AJC}}(g),
    \quad
    \theta\in\RR.
\end{align}
Indeed, $N_A=(\sigma_x\otimes\one)N_T(\sigma_x\otimes\one)$, consistently with the unitary equivalence \eqref{eq:JC/AJC_unitary}.

\subsection{Anisotropic Rabi model}\label{sec:anisotropic}
Throughout this paper, we identify $\CC^2 \otimes L^2(\RR) \cong L^2(\RR) \oplus L^2(\RR) \cong L^2(\RR; \CC^2)$ via 
\begin{align*}
    \binom{1}{0} \otimes f + \binom{0}{1} \otimes g 
    \cong 
    f \oplus g \cong \binom{f}{g}.
\end{align*} 
In what follows, we represent an operator
\begin{align*}
U:L^2(\mathbb{R})\oplus L^2(\mathbb{R})
 \longrightarrow L^2(\mathbb{R})\oplus L^2(\mathbb{R}),
\qquad
U\binom{f}{g}
=
\binom{Af+Bg}{Cf+Dg},
\end{align*}
in matrix form as
\begin{align*}
U=
\begin{pmatrix}
A & B\\
C & D
\end{pmatrix}.
\end{align*}
%or $U = T \oplus S$. 
With this matrix representation, the Hamiltonians are represented by
\begin{align*}
    &H_\mathrm{QR}(g) 
    = \begin{pmatrix} 
        a^\dagger a + \frac12 + \triangle & 0 \\
        0 & a^\dagger a + \frac12 - \triangle
    \end{pmatrix}
    + g \begin{pmatrix}
        0  & a + a^\dagger \\ 
        a + a^\dagger & 0
    \end{pmatrix}, \\
    &H_\mathrm{JC} (g) 
    = \begin{pmatrix}
        a^\dagger a + \frac12 + \triangle & 0 \\
        0 & a^\dagger a + \frac12 - \triangle
    \end{pmatrix}
    + g \begin{pmatrix} 
        0 & a \\ 
        a^\dagger & 0 
    \end{pmatrix}, \\
    &H_\mathrm{AJC}(g) 
    = \begin{pmatrix}
        a^\dagger a + \frac12 + \triangle & 0 \\
        0 & a^\dagger a + \frac12 - \triangle
    \end{pmatrix}
    + g \begin{pmatrix} 
        0 & a^\dagger \\ 
        a & 0 
    \end{pmatrix}.
\end{align*}
We generalize the quantum Rabi Hamiltonian and JC Hamiltonian mentioned above.  
An anisotropic Rabi Hamiltonian is defined for parameters $\lambda_1, \lambda_2 \in \RR$ by
\begin{align}\label{def:H_lambda1_lambda2}
    H_{\lambda_1, \lambda_2}(g) 
    = \begin{pmatrix} 
        a^\dagger a + \frac12 + \triangle & 0 \\
        0 & a^\dagger a + \frac12 - \triangle
    \end{pmatrix} 
    + g \begin{pmatrix}
        0 & \lambda_1 a + \lambda_2 a^\dagger \\
        \lambda_1 a^\dagger + \lambda_2 a & 0
    \end{pmatrix}. 
\end{align}
The case $\lambda_1 = \lambda_2$ reduces to the quantum Rabi model, whereas unequal coupling strengths $\lambda_1 \ne \lambda_2$ correspond to the anisotropic case. Expressing \eqref{def:H_lambda1_lambda2} in terms of the position operator $q$ 
and the momentum operator $p$ via 
\begin{align*}
    \alpha = \frac{\lambda_1 + \lambda_2}{\sqrt2},
    \quad
    \beta = \frac{\lambda_1 - \lambda_2}{\sqrt2},
\end{align*}
we define
\begin{align}\label{def:K_alpha_beta}
    K_{\alpha, \beta}(g) 
= \begin{pmatrix}
        h + \triangle & 0 \\
        0 & h - \triangle
    \end{pmatrix}
    + g \begin{pmatrix}
        0 & \alpha q + i\beta p \\
        \alpha q - i\beta p & 0
    \end{pmatrix}.
\end{align}
We also set 
\begin{align}\label{hh}
    H 
    = 
    \begin{pmatrix}
        h & 0\\
        0 & h
    \end{pmatrix}. 
    \end{align}
Note that 
\[K_{\alpha, \beta}(g)= H_{\frac{\alpha + \beta}{\sqrt2}, \frac{\alpha - \beta}{\sqrt2}}(g)\] 
and 
special cases are parameterized by:
\begin{align*}
    K_{\alpha,0}(g) = H_\mathrm{QR}\left(\frac{g\alpha}{\sqrt2}\right), 
    \quad
    K_{\alpha,\alpha}(g) = H_\mathrm{JC} (g\sqrt2 \alpha), 
    \quad
    K_{\alpha,-\alpha}(g) = H_\mathrm{AJC}(g\sqrt2 \alpha).
\end{align*}

\begin{lemma}\label{lem:sa}
For any $\alpha, \beta, g \in \RR$, $K_{\alpha,\beta}(g)$ is self-adjoint on $D(H)$ and essentially self-adjoint on any core for $H$. Moreover, $K_{\alpha,\beta}(g)$ is bounded from below and has a compact resolvent. In particular, its spectrum consists of discrete spectrum accumulating at $+\infty$.
\end{lemma}

\begin{proof}
For any $\eps>0$, standard estimates  yield $b_\eps>0$ such that 
$\|pf\|\le\eps\|hf\|+b_\eps\|f\|$ and $\|qf\|\le\eps\|hf\|+b_\eps\|f\|$. Consequently, the interaction matrix term in \eqref{def:K_alpha_beta} is infinitely $H_0$-small, where $H_0=\triangle\sigma_z\otimes\one+\one\otimes h$. Self-adjointness and essential self-adjointness then follow directly from the Kato-Rellich theorem~\cite{Kato51}. 

To establish the lower boundedness, we observe the identity:
\begin{align*}
    K_{\alpha, \beta}(g) 
    &= 
    \begin{pmatrix} \triangle & 0\\ 0 & -\triangle \end{pmatrix}
    + 
    \frac12 \begin{pmatrix} q & g\alpha\\ g\alpha & q \end{pmatrix}^2
    + 
    \frac12 \begin{pmatrix} p & ig\beta\\ -ig\beta & p \end{pmatrix}^2
    - 
    \frac{g^2 (\alpha^2 + \beta^2)}{2}.
\end{align*}
Since the quadratic terms are non-negative self-adjoint operators, we immediately obtain the explicit lower bound:
\begin{align}\label{eq:lower_bound}
    K_{\alpha,\beta}(g) 
    \ge 
    -|\triangle| - \frac{g^2 (\alpha^2 + \beta^2)}{2} 
    > -\infty.
\end{align}

Compactness of the resolvent follows, since $H_0$ has a compact resolvent and the interaction term is a relatively compact perturbation.
\end{proof}
\begin{remark}\label{rem:renormalization}
The lower bound \eqref{eq:lower_bound}  does not determine the leading-order behavior of the lowest eigenvalues as $|g|\to\infty$. As shown in this paper, the appropriate quadratic renormalization is
given by 
\begin{align*}
    \frac{g^2 \max\{\alpha^2,\beta^2\}}{2}. 
\end{align*}
After this renormalization, we identify the limiting operator governing the strong-coupling asymptotics.
\end{remark}
\begin{lemma}\label{lem:symm}
For any  $\alpha,\beta,g\in\RR$, $K_{\alpha,\beta}(g)$ possesses the $\ZZ_2$-symmetry generated by $P$. In addition,
\begin{align*}
    e^{i\theta N_T} K_{\alpha,\beta}(g) e^{-i\theta N_T}
    = 
    K_{\alpha,\beta}(g),
    \quad 
\theta\in\RR
\end{align*}
if and only if $g(\alpha-\beta)=0$, whereas
\begin{align*}
    e^{i\theta N_A} K_{\alpha,\beta}(g) e^{-i\theta N_A}
    = 
    K_{\alpha,\beta}(g),
    \quad 
\theta\in\RR
\end{align*}
if and only if $g(\alpha+\beta)=0$.
\end{lemma}
\begin{proof}
Since
\begin{align*}
\Pi q\Pi = -q,
\quad    
\Pi p\Pi = -p,
\quad    
\sigma_z\sigma_x\sigma_z = -\sigma_x,
\quad
\sigma_z\sigma_y\sigma_z = -\sigma_y,
\end{align*}
we have
$PK_{\alpha,\beta}(g)P=K_{\alpha,\beta}(g)$,
which proves the $\ZZ_2$-symmetry.
The interaction term can be written as
\begin{align*}
    \frac{g(\alpha+\beta)}{\sqrt{2}}
    \left(\sigma_+\otimes a+\sigma_-\otimes a^\dagger\right)
    +\frac{g(\alpha-\beta)}{\sqrt{2}}
    \left(\sigma_+\otimes a^\dagger+\sigma_-\otimes a\right).
\end{align*}
The first term is invariant under conjugation by $e^{i\theta N_T}$,
whereas the two summands in the second term acquire the phases
$e^{\pm 2i\theta}$. Hence, $K_{\alpha,\beta}(g)$ possesses the
$U(1)$-symmetry generated by $N_T$ if and only if
$g(\alpha-\beta)=0$.
Similarly, the second term is invariant under conjugation by
$e^{i\theta N_A}$, whereas the two summands in the first term acquire
the phases $e^{\mp 2i\theta}$. Thus, $K_{\alpha,\beta}(g)$ possesses
the $U(1)$-symmetry generated by $N_A$ if and only if
$g(\alpha+\beta)=0$.\end{proof}
\begin{lemma}\label{lem:exchange}
For any $\alpha, \beta, g \in \RR$, $K_{\alpha,\beta}(g)$ is unitarily equivalent to $K_{\beta,\alpha}(g)$.
\end{lemma}
\begin{proof}
Let $u = e^{-i\frac{\pi}{4}\sigma_z}$ and let $F = e^{-i\frac{\pi}{2}a^\dagger a}$ be the Fourier transform on $L^2(\RR)$. Then $u \sigma_x u^{-1} = \sigma_y$, $u \sigma_y u^{-1} = -\sigma_x$, $FqF^{-1} = -p$, and $FpF^{-1} = q$. Defining the unitary operator $\mathcal{F} = u \otimes F$ on $\CC^2 \otimes L^2(\RR)$, we find $\mathcal{F} K_{\alpha,\beta}(g) \mathcal{F}^{-1} = K_{\beta,\alpha}(g)$.
\end{proof}
We have the corollary. 
\begin{corollary}
$H_{\lambda,\lambda}(g)$ and  
$H_{\lambda,-\lambda}(g)$ are unitarily equivalent for any $\lambda\in\RR$.
\end{corollary}
\medskip
Having established these foundational properties, our goal in Section~3 is to analyze the asymptotic behavior of $K_{\alpha,\beta}(g)$ as $|g|\to\infty$. We assume throughout that $(\alpha,\beta)\ne(0,0)$. 
\section{Strong-coupling asymptotics}\label{s3}
This section is devoted to studying the strong-coupling limit of $K_{\alpha,\beta}(g)$.
By the unitary equivalence in Lemma~\ref{lem:exchange}, it is sufficient  to consider the regime $|\alpha|\ge|\beta|$ with $\alpha\ne0$. We first prove strong resolvent convergence of a suitably transformed and renormalized Hamiltonian, including the critical case $|\alpha|=|\beta|$. Away from the critical lines, we then improve the convergence to norm resolvent convergence. The regime $|\alpha|<|\beta|$ is then obtained by the unitary equivalence in Lemma~\ref{lem:exchange}. Finally, we reduce the Hamiltonian to the eigenspace of the parity operator $P$  and derive the asymptotic behavior of the eigenvalues in each eigenspace.
\subsection{Unitary transformations and renormalization}
Assume $\alpha\ne0$ and $|\alpha|\ge|\beta|$. We apply a unitary transformation to $K_{\alpha,\beta}(g)$ to reveal the renormalization constant. First, by applying a unitary transformation to the spin part, we see that
\begin{align*}
    e^{i\frac{\pi}{4}\sigma_y} K_{\alpha,\beta}(g) e^{-i\frac{\pi}{4}\sigma_y}
    = 
    \begin{pmatrix}
        \frac12 (p^2 + (q + g\alpha)^2) & -\triangle + ig\beta p\\
        -\triangle - ig\beta p & \frac12 (p^2 + (q - g\alpha)^2)
    \end{pmatrix} 
    - \frac{g^2\alpha^2}{2}.
\end{align*}
So we define the renormalized Hamiltonian by 
\begin{align*}
    K^\mathrm{ren} _{\alpha,\beta}(g) = K_{\alpha,\beta}(g) + \frac{g^2\alpha^2}{2}
\end{align*}
and transform it by the unitary
\[
    \mathcal{U} _{g\alpha} 
    = 
    \begin{pmatrix} e^{-ig\alpha p} & 0\\ 0 & e^{ig\alpha p} \end{pmatrix}
    e^{i\frac{\pi}{4}\sigma_y}
    = 
    e^{-ig\alpha \sigma_z \otimes p} e^{i\frac{\pi}{4}\sigma_y},\]
then we have
\begin{align*}
    \hat{K}^{\mathrm{ren} }_{\alpha,\beta}(g)
    = 
    \mathcal{U} _{g\alpha} 
    K^\mathrm{ren} _{\alpha,\beta}(g) 
    \mathcal{U} _{g\alpha}^{-1} 
    = 
    \begin{pmatrix}
        h & e^{-2ig\alpha p} (-\triangle + ig\beta p)\\
        e^{2ig\alpha p} (-\triangle - ig\beta p) & h 
    \end{pmatrix}.
\end{align*}
We study the asymptotic behavior of the resolvent $(\hat{K}^{\mathrm{ren} }_{\alpha,\beta}(g) - z)^{-1}$ as $|g|\to\infty$. %This approach allows us to extract information on the asymptotic spectral structure of $K_{\alpha,\beta}(g)$.
We first recall the well-known result in the isotropic case $\beta=0$. 
Let
\begin{align*}
    V_g 
    = 
    \begin{pmatrix}
        0 & -\triangle e^{-2ig\alpha p}\\
        -\triangle e^{2ig\alpha p} & 0
    \end{pmatrix}. 
\end{align*}
Then we have 
$
    \hat{K}^\mathrm{ren} _{\alpha,0}(g) = H + V_g
$. 
In this case, $V_g$ is a bounded operator with $\|V_g\|=|\triangle|$, and it converges weakly to 0 as $|g|\to\infty$. The following result is well known.
\begin{proposition}[\cite{Hirokawa15, HMS17}]\label{prop:iso}
Suppose that $|\alpha|>|\beta|=0$. Then 
\begin{align*}
    \hat{K}^\mathrm{ren} _{\alpha,0}(g) \longrightarrow 
    H\end{align*}
as $|g|\to\infty$ in the norm resolvent sense. 
%In other words, $\hat{K}^\mathrm{ren} _{\alpha,0}(g)$ converges to $H$ in the norm resolvent sense.
\end{proposition}
\begin{proof}
Let $z\in\CC\setminus\RR$. 
Using the second resolvent identity twice yields that 
\begin{align*}
    &\left(H + V_g - z\right)^{-1} - \left(H - z\right)^{-1}
%    \\ &=     -\left(H + V_g - z\right)^{-1} V_g \left(H - z\right)^{-1}
    \\
    &= 
    -\left(H - z\right)^{-1} V_g \left(H - z\right)^{-1} 
    + 
    \left(H + V_g - z\right)^{-1} V_g \left(H - z\right)^{-1} V_g \left(H - z\right)^{-1}.
\end{align*}
Since $(H-z)^{-1}$ is compact and $V_g$ weakly converges to 0 as $|g|\to\infty$, we see that the first term goes to 0 in the operator norm. In addition, since $\| (H + V_g - z)^{-1} V_g \| \leq |\triangle| / |\Im z|$, we conclude that the second term also goes to 0 in the operator norm.
\end{proof}
When $\beta\ne0$, however, the terms $\pm ig\beta p e^{\mp2ig\alpha p}$ in the off-diagonal parts of $\hat{K}^\mathrm{ren} _{\alpha,\beta}(g)$ contain the unbounded operator $p$ together with the growing parameter $g$. Consequently, the argument used in the isotropic case cannot be applied directly. To address this, we exploit the matrix structure of the operator and analyze its resolvent via the Schur complement.
\subsection{Effective harmonic oscillators and strong resolvent convergences}% via Schur complements}
We present an abstract theory of Schur complements for operator matrices. Let \begin{align*}M = \begin{pmatrix} A & B\\ C & D \end{pmatrix}:L^2(\RR)\oplus L^2(\RR)\to L^2(\RR)\oplus L^2(\RR),\end{align*} where $A, B, C$ and $D$ are linear operators in $L^2(\RR)$. 
Suppose first that $A$ is invertible. 
Then 
\begin{align*}
    T = D - CA^{-1}B
\end{align*} is called the Schur complement of $A$ in $M$. 
Formally $M$ admits the LDU factorization: 
\begin{align*}
    M
    =
    \begin{pmatrix}
        1 & 0\\
        CA^{-1} & 1
    \end{pmatrix}
    \begin{pmatrix}
        A & 0\\
        0 & T
    \end{pmatrix}
    \begin{pmatrix}
        1 & A^{-1}B\\
        0 & 1
    \end{pmatrix}.
\end{align*}
Similarly, suppose next that $D$ is invertible. Then $M$ formally admits the UDL factorization:
\begin{align*}
    M
    =
    \begin{pmatrix}
        1 & BD^{-1}\\
        0 & 1
    \end{pmatrix}
    \begin{pmatrix}
        S & 0\\
        0 & D
    \end{pmatrix}
    \begin{pmatrix}
        1 & 0\\
        D^{-1}C & 1
    \end{pmatrix}.
\end{align*}
The operator
\begin{align*}
    S = A - BD^{-1}C
\end{align*}
is also called the Schur complement of $D$ in $M$. We shall employ these factorizations to investigate the limit of the resolvent of $\hat{K}^\mathrm{ren} _{\alpha,\beta}(g)$. 

Let \begin{align*}
    X_g = e^{-2ig\alpha p}(-\triangle + ig\beta p).
\end{align*}
Then
\begin{align*}
    \hat{K}^\mathrm{ren} _{\alpha,\beta}(g) 
    = 
    \begin{pmatrix}
        h & X_g\\
        X_g^\ast & h 
    \end{pmatrix}.
\end{align*}
For $z\in\CC\setminus\RR$, define the Schur complements
\begin{align*}
    &T_g(z) = h - z - X_g^\ast (h - z)^{-1} X_g,\\
    &S_g(z) = h - z - X_g (h - z)^{-1} X_g^\ast. 
    \end{align*}
We first derive a matrix representation of the resolvent in terms of these operators.
\begin{lemma}\label{lem:block_rep}
Let $z\in\CC\setminus\RR$. Then $S_g(z)$ and $T_g(z)$ are invertible. 
Moreover, we have the matrix representation: 
\begin{align}\label{block}
    \begin{pmatrix}
        h - z & X_g\\
        X_g^\ast & h - z
    \end{pmatrix}^{-1}
    =
    \begin{pmatrix}
        A_g(z) & B_g(z)\\
        C_g(z) & D_g(z)
    \end{pmatrix},
\end{align}
where the matrix entries are given by
\begin{align*}
    A_g(z) &= (h - z - X_g (h - z)^{-1} X_g^\ast)^{-1}
    \\
    &= (h - z)^{-1} + (h - z)^{-1} X_g D_g(z) X_g^\ast (h - z)^{-1},
    \\
    B_g(z) &= -A_g(z) X_g (h - z)^{-1}
    \\
    &= -(h - z)^{-1} X_g D_g(z),
    \\
    C_g(z) &= -(h - z)^{-1} X_g^\ast A_g(z)
    \\
    &= -D_g(z) X_g^\ast (h - z)^{-1},
    \\
    D_g(z) &= (h - z - X_g^\ast (h - z)^{-1} X_g)^{-1}
    \\
    &= (h - z)^{-1} + (h - z)^{-1} X_g^\ast A_g(z) X_g (h - z)^{-1}.
\end{align*}
\end{lemma}
\begin{proof}
For $\psi \in D(S_g(z))$, we have
\begin{align*}
    \Im (\psi, S_g(z)\psi) 
    &= 
    -(\Im z) \| \psi \|^2 - \Im (X_g^\ast \psi, (h - z)^{-1} X_g^\ast \psi) 
    \\
    &= 
    -(\Im z) \| \psi \|^2 - (X_g^\ast \psi, (h - \bar{z})^{-1}(\Im z)(h-z)^{-1} X_g^\ast \psi) 
    \\
    &= 
    -(\Im z) \| \psi \|^2 - (\Im z) \| (h - z)^{-1} X_g^\ast \psi \|^2.
\end{align*}
This implies $|\Im (\psi, S_g(z)\psi)| \geq |\Im z| \| \psi \|^2$, which yields
\begin{align}\label{inj}
    \| S_g(z) \psi \| \geq |\Im z| \| \psi \|.
\end{align}
By \eqref{inj}, $S_g(z)$ is injective. Moreover, since $X_g (h - z)^{-1} X_g^\ast$ is bounded, $S_g(z)$ is a closed operator. This fact, combined with \eqref{inj}, yields the closedness of $\Ran(S_g(z))$. Note that $S_g(z)^\ast = S_g(\bar{z})$. Thus, we have the orthogonal decomposition
\begin{align*}
    L^2(\RR)
    =
    \overline{\Ran(S_g(z))} \oplus \Ker(S_g(z)^\ast)
    =
    \Ran(S_g(z)) \oplus \Ker(S_g(\bar{z})).
\end{align*}
Since $S_g(\bar{z})$ is injective, $S_g(z)$ is surjective. Therefore, $S_g(z)$ is bijective as a map from $D(S_g(z))$ to $L^2(\RR)$, and its inverse $S_g(z)^{-1}$ is a bounded operator satisfying $\| S_g(z)^{-1} \| \leq 1 / |\Im z|$.
The same argument, with $X_g$ and $X_g^\ast$ interchanged, shows that $T_g(z)$ is invertible. The UDL and LDU factorizations
\begin{align*}
    \begin{pmatrix}
        h - z & X_g\\
        X_g^\ast & h - z
    \end{pmatrix}
    &=
    \begin{pmatrix}
        1 & X_g (h - z)^{-1}\\
        0 & 1
    \end{pmatrix}
    \begin{pmatrix}
        S_g(z) & 0\\
        0 & h - z
    \end{pmatrix}
    \begin{pmatrix}
        1 & 0\\
        (h - z)^{-1} X_g^\ast & 1
    \end{pmatrix}
    \\
    &=
    \begin{pmatrix}
        1 & 0\\
        X_g^\ast (h - z)^{-1} & 1
    \end{pmatrix}
    \begin{pmatrix}
        h - z & 0\\
        0 & T_g(z)
    \end{pmatrix}
    \begin{pmatrix}
        1 & (h - z)^{-1} X_g\\
        0 & 1
    \end{pmatrix}
\end{align*}
give the asserted identities for the matrix entries.
\end{proof}
Next, we consider the strong-coupling limit of each matrix entry in \eqref{block}. Since $\hat{K}^\mathrm{ren} _{\alpha,\beta}(g)$ is self-adjoint, the standard resolvent estimate gives $\| (\hat{K}^\mathrm{ren} _{\alpha,\beta}(g) - z)^{-1} \| \leq 1 / |\Im z|$. Consequently, all matrix entries are uniformly bounded in $g$:
\begin{align*}
    \max\{ \| A_g(z) \|, \| B_g(z) \|, \| C_g(z) \|, \| D_g(z) \| \} 
    \leq \frac{1}{ |\Im z|}.
\end{align*}
The harmonic oscillator $h$ and its direct sum $H$ are defined by \eqref{h} and \eqref{hh}, respectively. 
\begin{definition}[Effective harmonic oscillators]\label{eff}
We define the effective harmonic oscillators $\hat{h}_\infty$ and $\check{h}_\infty$  by
\begin{align*}
&\hat{h}_\infty
=
\frac12\left(
q^2+\left(1-\frac{\beta^2}{\alpha^2}\right)p^2
\right),\\
&\check{h}_\infty
=
\frac12\left(
p^2+\left(1-\frac{\alpha^2}{\beta^2}\right)q^2
\right),
\end{align*}
and we set
\begin{align*}
&\hat{H}_\infty
=
\begin{pmatrix}
\hat{h}_\infty & 0 \\
0 & \hat{h}_\infty
\end{pmatrix},\\
&\check{H}_\infty
=
\begin{pmatrix}
\check{h}_\infty & 0 \\
0 & \check{h}_\infty
\end{pmatrix}.
\end{align*}
\end{definition}
For $|\alpha|>|\beta|$, it follows that %$\hat{h}_\infty$ is a harmonic oscillator with
\begin{align*}
    \operatorname{spec}(\hat{h}_\infty)
    =
    \left\{
        \sqrt{1 - \frac{\beta^2}{\alpha^2}}
        \left(n + \frac12\right)
    \right\}_{n=0}^\infty
\end{align*}
and 
for $|\alpha|<|\beta|$, %it follows that %$\hat{h}_\infty$ is a harmonic oscillator with
\begin{align*}
    \operatorname{spec}(\check{h}_\infty)
    =
    \left\{
        \sqrt{1 - \frac{\alpha^2}{\beta^2}}
        \left(n + \frac12\right)
    \right\}_{n=0}^\infty.
\end{align*}
At the critical value $|\alpha|=|\beta|$, it reduces to $q^2/2$ and 
\[\spec(\hat{h}_\infty) = \spec(\check{h}_\infty)=[0,\infty).\] Let $\mathscr{S}(\RR)$ denote the Schwartz space of rapidly decreasing smooth functions on $\RR$. The next lemma is a key lemma of this paper.
\begin{lemma}\label{lem:diag_conv}
Assume $\alpha\ne0$ and $|\alpha|\ge|\beta|$. Let $z\in\CC\setminus\RR$. 
Then 
\begin{align*}
&\operatorname*{s-lim}_{|g|\to\infty} A_g(z)=(\hat{h}_\infty-z)^{-1},\\
&\operatorname*{s-lim}_{|g|\to\infty} D_g(z)=(\hat{h}_\infty-z)^{-1}.
\end{align*}
\end{lemma}
\begin{proof}
By symmetry, it suffices to prove the strong convergence for $A_g(z)$, as the proof for $D_g(z)$ is identical upon replacing $X_g$ with $X_g^\ast$.
Since $\sup_{g} \| A_g(z) \| < \infty$, ${\mathscr S}(\RR) \subset \bigcap_g D(A_g(z)^{-1}) \cap D(\hat{h}_\infty)$, and ${\mathscr S}(\RR)$ is a core for $h_{\infty} - z$, it suffices to establish the following limit on ${\mathscr S}(\RR)$:
\begin{align*}
    \lim_{|g|\to\infty} X_g (h - z)^{-1} X_g^\ast \psi 
    = \frac{\beta^2}{2\alpha^2} p^2 \psi, 
    \quad \psi \in {\mathscr S}(\RR).
\end{align*}
For any $\psi \in {\mathscr S}(\RR)$, we can write
\begin{align*}
    X_g (h - z)^{-1} X_g^\ast \psi
    = (-\triangle + ig\beta p) (h_g - z)^{-1} (-\triangle - ig\beta p) \psi,
\end{align*}
where 
\begin{align}h_g = e^{-2ig\alpha p} h e^{2ig\alpha p} = \frac12 (p^2 + (q - 2g\alpha)^2) = h - 2g\alpha q + 2g^2\alpha^2.\end{align} 
Thus our task reduces to evaluating the asymptotic behavior of
\begin{align*}
    g^n (h_g - z)^{-1}\psi, \quad g^n p (h_g - z)^{-1} \psi \quad (n = 0, 1, 2)
\end{align*}
as $|g| \to \infty$ for $\psi \in {\mathscr S}(\RR)$. Applying the second resolvent identity, we have
\begin{align*}
    \left( (h_g - z)^{-1} - \frac{1}{2g^2\alpha^2} \right) \psi
    = -(h_g - z)^{-1} (h - z - 2g\alpha q) \frac{1}{2g^2\alpha^2} \psi,
\end{align*}
which yields the identity
\begin{align}\label{n=0eq}
    (h_g - z)^{-1} \psi
    = \frac{1}{2g^2\alpha^2} \psi 
    - \frac{1}{2g^2\alpha^2} (h_g - z)^{-1} (h - z) \psi
    + \frac{1}{g\alpha} (h_g - z)^{-1} q\psi.
\end{align}
Since $\| (h_g - z)^{-1} \| = \| (h - z)^{-1} \| < \infty$, identity \eqref{n=0eq} immediately implies
\begin{align}\label{n=0lim}
    \lim_{|g|\to\infty} (h_g - z)^{-1} \psi = 0, 
    \quad 
    \psi \in {\mathscr S}(\RR).
\end{align}
Multiplying both sides of \eqref{n=0eq} by $g$ and combining with \eqref{n=0lim}, we obtain
\begin{align}\label{n=1lim}
    \lim_{|g|\to\infty} g (h_g - z)^{-1} \psi = 0, 
    \quad 
    \psi \in {\mathscr S}(\RR).
\end{align}
Furthermore, multiplying both sides of \eqref{n=0eq} by $g^2$ and using \eqref{n=0lim} and \eqref{n=1lim}, we get
\begin{align*}
    \lim_{|g|\to\infty} g^2 (h_g - z)^{-1} \psi = \frac{1}{2\alpha^2} \psi, 
    \quad 
    \psi \in {\mathscr S}(\RR).
\end{align*}
We emphasize that $g^n (h_g - z)^{-1}, n = 1, 2$ cannot converge strongly on the whole $L^2(\RR)$ since they lack uniform boundedness. We can only obtain strong convergence on a proper subspace. Next, multiplying $p$ from the left onto both sides of \eqref{n=0eq}, we have
\begin{align*}
    p (h_g - z)^{-1} \psi
    = \frac{1}{2g^2\alpha^2} p\psi 
    - \frac{1}{2g^2\alpha^2} p (h_g - z)^{-1} (h - z) \psi
    + \frac{1}{g\alpha} p (h_g - z)^{-1} q\psi.
\end{align*}
Since $\| p (h_g - z)^{-1} \| = \| p (h - z)^{-1} \| < \infty$, a similar argument yields that
\begin{align*}
    \lim_{|g|\to\infty} p (h_g - z)^{-1} \psi = \lim_{|g|\to\infty} g p (h_g - z)^{-1} \psi = 0, 
    \quad \text{and} \quad
    \lim_{|g|\to\infty} g^2 p (h_g - z)^{-1} \psi = \frac{1}{2\alpha^2} p\psi
\end{align*}
for all $\psi \in {\mathscr S}(\RR)$. Combining these asymptotic limits completes the proof for $A_g(z)$.
\end{proof}
\begin{lemma}\label{lem:offdiag_conv}
Assume $\alpha\ne0$ and $|\alpha|\ge|\beta|$. Let $z\in\CC\setminus\RR$. Then 
Then 
\begin{align*}
&\operatorname*{s-lim}_{|g|\to\infty} B_g(z)=0,\\
&\operatorname*{s-lim}_{|g|\to\infty} C_g(z)=0.
\end{align*}
\end{lemma}
\begin{proof}
Set
\begin{align*}
    R_g(z)
    =
    (\hat{K}_{\alpha,\beta}^{\mathrm{ren} }(g) - z)^{-1}
    =
    \begin{pmatrix}
        A_g(z) & B_g(z)\\
        C_g(z) & D_g(z)
    \end{pmatrix}.
\end{align*}
Note that
\begin{align*}
    R_g(\bar z) = R_g(z)^*,
    \quad
    R_g(z) - R_g(\bar z) = (z - \bar z) R_g(\bar z) R_g(z).
\end{align*}
Hence, for every $\Psi \in L^2(\RR) \oplus L^2(\RR)$,
\begin{align}\label{eq:resolvent-imaginary-part}
    \Im (\Psi, R_g(z)\Psi)
    =
    (\Im z) \| R_g(z)\Psi \|^2.
\end{align}
Let $\psi \in L^2(\RR)$. Taking $\Psi = \binom{\psi}{0}$ in \eqref{eq:resolvent-imaginary-part} gives
\begin{align*}
    \Im(\psi, A_g(z)\psi)
    =
    (\Im z)
    \left(
        \| A_g(z)\psi \|^2 + \| C_g(z)\psi \|^2
    \right).
\end{align*}
By Lemma~\ref{lem:diag_conv}, we have
\begin{align*}
    \lim_{|g|\to\infty} \| A_g(z)\psi \|^2
    =
    \| (\hat{h}_\infty - z)^{-1}\psi \|^2
\end{align*}
and
\begin{align*}
    \lim_{|g|\to\infty} \Im(\psi, A_g(z)\psi)
    =
    \Im\left(\psi, (\hat{h}_\infty - z)^{-1}\psi\right)
    =
    (\Im z) \| (\hat{h}_\infty - z)^{-1}\psi \|^2.
\end{align*}
Since $\Im z\ne0$, it follows that $\lim_{|g|\to\infty}\|C_g(z)\psi\| = 0$.
Similarly, taking $\Psi = \binom{0}{\psi}$ in \eqref{eq:resolvent-imaginary-part} yields
\begin{align*}
    \Im(\psi, D_g(z)\psi)
    =
    (\Im z)
    \left(
        \| B_g(z)\psi \|^2 + \| D_g(z)\psi \|^2
    \right).
\end{align*}
The strong convergence $\lim_{|g|\to\infty}D_g(z) = (\hat{h}_\infty - z)^{-1}$ from Lemma~\ref{lem:diag_conv} therefore implies $\lim_{|g|\to\infty}\|B_g(z)\psi\| = 0$.
\end{proof}
Combining the convergence of the diagonal and off-diagonal entries of the resolvent, we obtain the strong resolvent limit of the full operator. 
\begin{lemma}\label{prop:strong_resolvent_conv}
Suppose that $\alpha\ne0$ and $|\alpha|\ge|\beta|$. Then for all $z\in\CC\setminus\RR$, 
\begin{align*}
\operatorname*{s-lim}_{|g|\to\infty}
    (\hat{K}_{\alpha,\beta}^{\mathrm{ren} }(g) - z)^{-1}
    =
    (\hat{H}_\infty - z)^{-1}.
\end{align*}
\end{lemma}
\begin{proof}
The assertion follows immediately from Lemmas~\ref{lem:block_rep}, \ref{lem:diag_conv}, and \ref{lem:offdiag_conv}.
\end{proof}
\subsection{Norm resolvent convergences}
We next show the norm resolvent convergence of $\hat{K}_{\alpha,\beta}^\mathrm{ren} $ 
whenever $|\alpha|\ne|\beta|$. 
The following abstract lemma is used to upgrade strong resolvent convergences to 
norm resolvent convergences.
\begin{lemma}[Compact-domination lemma]\label{lem:compact_domination}
Let $T_n$, $T$, and $K$ be bounded nonnegative self-adjoint operators on a Hilbert space. Assume that $K$ is compact,
$\operatorname*{s-lim}_{n\to\infty}T_n = T$ and 
$
    0 \le T_n \le K$ 
for every $n$. Then it follows that 
\begin{align*}\lim_{n\to\infty}\|T_n-T\|=0.\end{align*}
\end{lemma}
\begin{proof}
First, passing to the limit in quadratic forms gives $0\le T\le K$. Let $P$ be a finite-rank orthogonal projection. Since $\Ran P$ is finite-dimensional and $\operatorname*{s-lim}_{n\to\infty} T_n=T$,
\begin{align*}
    \lim_{n\to\infty} \| (T_n-T)P \| = 0.
\end{align*}
Since
$
    T_n^2 
    = 
    T_n^{1/2} T_n T_n^{1/2} 
    \le 
    \|T_n\| T_n
    \le
    \|K\| K$, 
we have
\begin{align*}
    \|T_n(1-P)\|^2
    =
    \|(1-P)T_n^2(1-P)\|
    \le
    \|K\|\,\|(1-P)K(1-P)\|.
\end{align*}
The same estimate holds with $T$ in place of $T_n$. 
Let $\epsilon>0$. 
Since $K$ is compact, we can choose finite-rank projections $P_\epsilon$ such that
$\|(1-P_\epsilon)K(1-P_\epsilon)\|\leq \epsilon$.  
Therefore
\begin{align*}
    \|T_n-T\|
    &\le
    \|(T_n-T)P_\epsilon\|+\|T_n(1-P_\epsilon)\|+\|T(1-P_\epsilon)\|\leq 2\|K\|\epsilon 
    +\|(T_n-T)P_\epsilon\|.
\end{align*}
We can see that $\lim_{n\to\infty}\|T_n-T\|\leq 2\|K\|\epsilon$. Then the lemma follows. 
\end{proof}

We shall show the norm resolvent convergence of $\hat{K}_{\alpha,\beta}^\mathrm{ren} $ by employing 
Lemma \ref{lem:compact_domination}. 
The essential input is to show the uniform lower bound of $\hat{K}_{\alpha,\beta}^{\mathrm{ren} }(g)$ with respect to $g$. 
\begin{lemma}\label{prop:uniform-lower-bound}
Suppose $|\alpha|>|\beta|$ and set
\begin{align}\label{c}
    \eta = 1 - \frac{|\beta|}{|\alpha|} > 0,
    \quad
    c = |\triangle| + \frac{|\beta|}{2|\alpha|}. 
\end{align}
Then, for every $g\in\RR$,
\begin{align}\label{eq:uniform-lower-bound}
    \hat{K}_{\alpha,\beta}^{\mathrm{ren} }(g)
    \ge
    \eta H - c.
\end{align}
\end{lemma}
\begin{proof}
Since
\begin{align}\label{eq:Rabi-lower-bound}
    \hat{K}_{\alpha,0}^{\mathrm{ren} }(g)
    =
    \begin{pmatrix}
        h & -\triangle e^{-2ig\alpha p}\\
        -\triangle e^{2ig\alpha p} & h
    \end{pmatrix}
    \ge 
    H - |\triangle|,
\end{align}
the assertion follows immediately when $\beta=0$.
Suppose now that $\beta\ne0$, and set $r=|\beta/\alpha|\in(0,1)$ and $s=\operatorname{sgn}(\beta/\alpha)\in\{\pm1\}$. Then $K_{\alpha,\beta}(g)$ admits the convex decomposition
\begin{align*}
    K_{\alpha,\beta}(g)
    =
    (1 - r) K_{\alpha,0}(g) + r K_{\alpha,s\alpha}(g).
\end{align*}
If the same quadratic renormalization $g^2\alpha^2/2$ and the same unitary transformation $\mathcal{U} _{g\alpha}$ are applied to all three operators, then we obtain
\begin{align}\label{eq:convex-decomposition}
    \hat{K}_{\alpha,\beta}^{\mathrm{ren} }(g)
    &=
    (1 - r) \hat{K}_{\alpha,0}^{\mathrm{ren} }(g) 
    + 
    r \hat{K}_{\alpha,s\alpha}^{\mathrm{ren} }(g).
\end{align}
We next estimate the endpoint $\hat{K}_{\alpha,s\alpha}^{\mathrm{ren} }(g)$.
For $s=1$, using the correspondence with the JC Hamiltonian,
we have
\begin{align*}
    K_{\alpha,\alpha}^{\mathrm{ren} }(g)
=
    \left(
        \begin{pmatrix} 0 & a\\ a^\dagger & 0 \end{pmatrix}
        + 
        \frac{g\alpha}{\sqrt2}
    \right)^2
    +
    \left(\triangle - \frac12\right)
    \begin{pmatrix} 1 & 0\\ 0 & -1 \end{pmatrix}.
\end{align*}
Therefore,
\begin{align*}
    \hat{K}_{\alpha,\alpha}^{\mathrm{ren} }(g)
    \cong
    K_{\alpha,\alpha}^{\mathrm{ren} }(g)
    \ge
    -\left| \triangle - \frac12 \right|
    \ge
    -|\triangle| - \frac12.
\end{align*}
For $s=-1$, by the unitary equivalence between the AJC Hamiltonian and the JC Hamiltonian with $\triangle$ replaced by $-\triangle$, as noted in \eqref{eq:JC/AJC_unitary}, the same estimate follows. Hence, for either $s=\pm1$,
\begin{align}\label{eq:JC/AJC-lower-bound}
    \hat{K}_{\alpha,s\alpha}^{\mathrm{ren} }(g)
    \ge
    -|\triangle| - \frac12.
\end{align}
Combining \eqref{eq:convex-decomposition}, \eqref{eq:Rabi-lower-bound}, and \eqref{eq:JC/AJC-lower-bound}, we obtain \eqref{eq:uniform-lower-bound}.
\end{proof}
For $|\alpha|<|\beta|$ we set 
\[
    \mathcal{V}_{g\beta}
    =
    e^{-ig\beta \sigma_z \otimes q} e^{-i\frac{\pi}{4}\sigma_x}\]
    and  
    \begin{align*}
&\check{K}_{\alpha,\beta}(g)=        \mathcal{V}_{g\beta} K_{\alpha,\beta}(g) \mathcal{V}_{g\beta}^{-1},\\
&\check{K}_{\alpha,\beta}^\mathrm{ren}(g) =\check{K}_{\alpha,\beta}(g)+\frac{g^2 \beta^2}{2}.
\end{align*}
The critical set
\begin{align*}
\bigl\{(\alpha,\beta)\in\mathbb{R}^2
      \mid |\alpha|=|\beta|\bigr\}
\end{align*}
separates the two norm-resolvent regimes, and the limiting operator on
this set is qualitatively different from those in the two regimes.
We now state the main theorem in this paper. \begin{theorem}\label{thm:resolvent_conv}
Assume $(\alpha,\beta)\ne(0,0)$.
\begin{enumerate}
\item[(1)] Suppose that $|\alpha|>|\beta|$. 
%Let $\mathcal{U} _{g\alpha}=e^{-ig\alpha \sigma_z \otimes p} e^{i\frac{\pi}{4}\sigma_y}.$
Then 
\begin{align*}
\hat{K}_{\alpha,\beta}^\mathrm{ren}(g) 
\longrightarrow 
\hat{H}_\infty
\end{align*}
as $|g|\to\infty$ in the norm resolvent sense.

\item[(2)] Suppose that $|\alpha|=|\beta|$. Then
\begin{align*}
\hat{K}_{\alpha,\beta}^\mathrm{ren}(g) 
\longrightarrow 
    \begin{pmatrix}
        \frac12 q^2 & 0\\
        0 & \frac12 q^2
    \end{pmatrix}
\end{align*}
as $|g|\to\infty$ in the strong resolvent sense, but not in the norm resolvent sense.

\item[(3)] Suppose that $|\alpha|<|\beta|$. 
Then 
\begin{align*}
\check{K}_{\alpha,\beta}^\mathrm{ren}(g) 
\longrightarrow 
\check{H}_\infty
\end{align*}
as $|g|\to\infty$ in the norm resolvent sense.
\end{enumerate}
\end{theorem}

\begin{proof}
(1) By Lemma~\ref{prop:uniform-lower-bound}, we have
$
    \hat{K}_{\alpha,\beta}^{\mathrm{ren} }(g) + c
    \geq
    \eta H > 0$,
and hence
\begin{align*}
    0 < 
    (\hat{K}_{\alpha,\beta}^{\mathrm{ren} }(g) + c)^{-1}
    \le
    (\eta H)^{-1}.
\end{align*}
In particular
$
    \sup_g \|(\hat{K}_{\alpha,\beta}^{\mathrm{ren} }(g) + c)^{-1}\|
    < \infty
$. 
We can extend 
Lemma~\ref{prop:strong_resolvent_conv} for $z=-c$, where $c$ is given by \eqref{c}. 
Thus,  
\begin{align}\label{eq:conv_at_-c}
\operatorname*{s-lim}_{|g|\to\infty}
      (\hat{K}_{\alpha,\beta}^{\mathrm{ren} }(g) + c)^{-1}
    =
    (\hat{H}_\infty + c)^{-1}. 
\end{align}
Combining this with the compactness of $(\eta H)^{-1}$, Lemma~\ref{lem:compact_domination} improves the convergence \eqref{eq:conv_at_-c} to the norm sense. By the same resolvent identity, this implies the norm resolvent convergence in (1).

(2) The strong resolvent convergence follows from Lemma~\ref{prop:strong_resolvent_conv}. For each $g$, the operator on the left-hand side has a compact resolvent by Lemma~\ref{lem:sa}, whereas the operator on the right-hand side does not. Since a norm limit of compact operators is compact, the convergence cannot be in the norm resolvent sense.

(3) We use the unitary equivalence in Lemma~\ref{lem:exchange}. Recall
$
    \mathcal{F}
    =
    e^{-i\frac{\pi}{4}\sigma_z}
    \otimes
    e^{-i\frac{\pi}{2}a^\dagger a},
$
so that
$
    \mathcal{F} K_{\beta,\alpha}(g) \mathcal{F}^{-1}
    =
    K_{\alpha,\beta}(g).
$
Since $|\beta|>|\alpha|$, part (1), applied to $K_{\beta,\alpha}(g)$, gives
\begin{align*}
        \mathcal{U} _{g\beta} K_{\beta,\alpha}(g) \mathcal{U} _{g\beta}^{-1}
        +
        \frac{g^2 \beta^2}{2}
\longrightarrow 
    \begin{pmatrix}
        \frac12 \left(q^2 + \left(1 - \frac{\alpha^2}{\beta^2}\right) p^2\right) & 0\\
        0 & \frac12 \left(q^2 + \left(1 - \frac{\alpha^2}{\beta^2}\right) p^2\right)
    \end{pmatrix}
\end{align*}
as   $|g|\to\infty$ 
in the norm resolvent sense. Moreover,
\begin{align*}
    \mathcal{F} \mathcal{U} _{g\beta} \mathcal{F}^{-1}
    &=
    (e^{-i\frac{\pi}{2}a^\dagger a} e^{-ig\beta \sigma_z \otimes p} e^{i\frac{\pi}{2}a^\dagger a})
    (e^{-i\frac{\pi}{4}\sigma_z} e^{i\frac{\pi}{4}\sigma_y} e^{i\frac{\pi}{4}\sigma_z})
    \\
    &=
    e^{-ig\beta \sigma_z \otimes q} e^{-i\frac{\pi}{4}\sigma_x}
    =
    \mathcal{V}_{g\beta}.
\end{align*}
Conjugating the above norm resolvent convergence by $\mathcal F$ therefore yields the assertion.
\end{proof}
\subsection{Reduction by $\ZZ_2$-symmetry}
Recall from Lemma~\ref{lem:symm} that the anisotropic Rabi Hamiltonian $K_{\alpha,\beta}(g)$ commutes with the parity operator
\begin{align*}
    P = \sigma_z \otimes \Pi = \begin{pmatrix} \Pi & 0\\ 0 & -\Pi \end{pmatrix},
\end{align*}
where $\Pi = (-1)^{a^\dagger a}$. 
According to $P$, the total Hilbert space
$\cH $ decomposes into the orthogonal sum:
\begin{align*}
    \cH  = \cH _+ \oplus \cH _-,
\end{align*}
where $\cH _\pm$ are the eigenspaces of $P$ corresponding to the eigenvalues $\pm 1$:
\begin{align*}
    \cH _\pm 
    = 
    \Ker(P \mp 1) 
    = 
    \left\{ 
        \begin{pmatrix} \psi_1\\ \psi_2 \end{pmatrix} \in \cH  
        \;\middle|\; 
        \Pi\psi_1 = \pm\psi_1, \;\; \Pi\psi_2 = \mp\psi_2 
    \right\}.
\end{align*}
We now reduce the spectral analysis of $K_{\alpha,\beta}(g)$ to the two parity sectors ${\cH _\pm}$.

In the case $|\alpha|\ge|\beta|$, we utilize the unitary transformation $\mathcal{U} _{g\alpha}$. The transformed parity operator is given by
    \begin{align*}
    \hat{P}
    =
    \mathcal{U} _{g\alpha} P \mathcal{U} _{g\alpha}^{-1}
    =
    -\sigma_x \otimes \Pi
    =
    \begin{pmatrix} 0 & -\Pi\\ -\Pi & 0 \end{pmatrix},
\end{align*}
which is independent of $g$. The unitary $\mathcal{U} _{g\alpha}$ maps $\cH _\pm$ onto
\begin{align*}
    \hat{\cH} _\pm
    =
    \Ker(\hat{P} \mp 1)
    =
    \left\{
        \begin{pmatrix} \psi\\ \mp\Pi\psi \end{pmatrix}
        \;\middle|\;
        \psi \in L^2(\RR)
    \right\}.
\end{align*}
Define the unitary operator
\begin{align*}
    \hat{W}_\pm: \hat{\cH} _\pm \longrightarrow L^2(\RR),
    \quad
    \hat{W}_\pm \begin{pmatrix} \psi\\ \mp\Pi\psi \end{pmatrix}
    =
    \sqrt2\psi
\end{align*}
and set
\begin{align*}
    \hat{K}_{\alpha,\beta,\pm}(g)
=
    \hat{W}_\pm
    \left(
        \mathcal{U} _{g\alpha} \upharpoonright_{\cH_\pm}
    \right)
    \left(
        K_{\alpha,\beta}(g)\upharpoonright_{\cH_\pm}
        +
        \frac{g^2\alpha^2}{2}
    \right)
    \left(
        \mathcal{U} _{g\alpha} \upharpoonright_{\cH_\pm}
    \right)^{-1}
    \hat{W}_\pm^{-1}.
\end{align*}
Since
$
    \hat{W}_\pm
    \left(
        \mathcal{U} _{g\alpha} \upharpoonright_{\cH_\pm}
    \right)
$
is unitary from $\cH_\pm$ onto $L^2(\RR)$, we have
\begin{align*}
    \spec\left(\hat{K}_{\alpha,\beta,\pm}(g)\right)
    =
    \spec
    \left(
        K_{\alpha,\beta}(g)\upharpoonright_{\cH_\pm}
        +
        \frac{g^2\alpha^2}{2}
    \right).
\end{align*}

The case $|\alpha|<|\beta|$ is treated in the same way. Under the unitary transformation
$
    \mathcal{V}_{g\beta}
$
the parity operator becomes
\begin{align*}
    \check{P}
    =
    \mathcal{V}_{g\beta} P \mathcal{V}_{g\beta}^{-1}
    =
    -\sigma_y \otimes \Pi
    =
    \begin{pmatrix} 0 & i\Pi\\ -i\Pi & 0 \end{pmatrix}
\end{align*}
and we set 
\begin{align*}
    \check{\cH}_\pm
    =
    \Ker(\check{P} \mp 1)
    =
    \left\{
        \begin{pmatrix} \psi\\ \mp i\Pi\psi \end{pmatrix}
        \;\middle|\;
        \psi \in L^2(\RR)
    \right\}.
\end{align*}
Let
\begin{align*}
    \check{W}_\pm: \check{\cH}_\pm \longrightarrow L^2(\RR),
    \quad
    \check{W}_\pm
    \begin{pmatrix} \psi\\ \mp i\Pi\psi \end{pmatrix}
    =
    \sqrt2\psi,
\end{align*}
and define
\begin{align*}
    \check{K}_{\alpha,\beta,\pm}(g)
=
    \check{W}_\pm
    \left(
        \mathcal V_{g\beta}\upharpoonright_{\cH _\pm}
    \right)
    \left(
        K_{\alpha,\beta}(g)\upharpoonright_{\cH _\pm}
        +
        \frac{g^2\beta^2}{2}
    \right)
    \left(
        \mathcal V_{g\beta}\upharpoonright_{\cH _\pm}
    \right)^{-1}
    (\check{W}_\pm)^{-1}.
\end{align*}
Since 
$
    \check{W}_\pm
    \left(
        \mathcal V_{g\beta}\upharpoonright_{\cH _\pm}
    \right)
$
is unitary from $\cH_\pm$ onto $L^2(\RR)$, we have
\begin{align*}
    \spec\left(\check{K}_{\alpha,\beta,\pm}(g)\right)
    =
    \spec
    \left(
        K_{\alpha,\beta}(g)\upharpoonright_{\cH _\pm}
        +
        \frac{g^2\beta^2}{2}
    \right).
\end{align*}
\begin{theorem}\label{thm:parity-reduction}
Assume that $(\alpha,\beta)\ne(0,0)$.
\begin{enumerate}
\item[(1)]
Suppose that $|\alpha|>|\beta|$. Then 
\begin{align*}
\hat{K}_{\alpha,\beta,\pm}(g)\longrightarrow
\hat{h}_\infty
\end{align*}
as $|g|\to\infty$ in the norm resolvent sense.

\item[(2)]
Suppose that $|\alpha|=|\beta|$. Then
\begin{align*}
  \hat{K}_{\alpha,\beta,\pm}(g)\longrightarrow   \frac12 q^2
\end{align*}
as $|g|\to\infty$
in the strong resolvent sense, but not in the norm resolvent sense.

\item[(3)]
Suppose that $|\alpha|<|\beta|$. Then
\begin{align*}
\check{K}_{\alpha,\beta,\pm}(g)\longrightarrow
\check{h}_\infty
\end{align*}
as $|g|\to\infty$ in the norm resolvent sense.
\end{enumerate}
\end{theorem}
\begin{proof}
Let $|\alpha|\ge|\beta|$.
Recall that $\hat{\cH} _\pm$ is independent of $g$ and reduces both
$\hat{K}_{\alpha,\beta}^{\mathrm{ren} }(g)$ and $\hat{H}_\infty$. Moreover,
\begin{align*}
    \hat{W}_\pm
    \left(
        \hat{H}_\infty \upharpoonright_{\hat{\cH} _\pm}
    \right)
    \hat{W}_\pm^{-1}
    =
    \hat{h}_\infty.
\end{align*}
Hence Theorem~\ref{thm:resolvent_conv}~(1) and~(2), restricted to
$\hat{\cH} _\pm$ and conjugated by $\hat{W}_\pm$, yield respectively the norm resolvent convergence in (1) and the strong resolvent convergence in (2). The failure of norm resolvent convergence in (2) follows from the same compactness argument as in the proof of Theorem~\ref{thm:resolvent_conv}~(2): each $\hat{K}_{\alpha,\beta,\pm}(g)$ has compact resolvent, whereas $g^2/2$ does not.
The proof for the case of 
$|\alpha|<|\beta|$ is similar. 
\end{proof}
\section{Eigenvalue asymptotics}\label{s4}
We first consider the case of $|\alpha|\ne|\beta|$, where the limiting operator is a harmonic oscillator and the norm resolvent convergence is available. 
    The spectra of the limiting operators $\hat{h}_\infty$ and $\check{h}_\infty $ are given by
\begin{align*}
    \operatorname{spec} \left(\hat{h}_\infty\right)
    &=
    \left\{
        \sqrt{1 - \frac{\beta^2}{\alpha^2}}
        \left(n + \frac12\right)
    \right\}_{n=0}^\infty,
    \quad  |\alpha| > |\beta|,
    \\
    \operatorname{spec} \left(\check{h}_\infty \right)
    &=
    \left\{
        \sqrt{1 - \frac{\alpha^2}{\beta^2}}
        \left(n + \frac12\right)
    \right\}_{n=0}^\infty,
    \quad  |\alpha| < |\beta|.
\end{align*}
Moreover, each eigenvalue is simple.
We present an abstract lemma for eigenvalues of positive compact self-adjoint operators. 
\begin{lemma}\label{ev}
Let $T$ and $S$ be positive compact self-adjoint operators on a Hilbert space $\mathcal K$. 
Let $\mu_n(T)$ and $\mu_n(S)$ denote their $n$-th eigenvalues, arranged in nonincreasing order and counted with multiplicity, where the eigenvalue sequences are extended by zeros if the operators have finite rank. Then, for every $n\geq 1$,
\[
    \left|\mu_n(T)-\mu_n(S)\right|
    \leq
    \|T-S\|.
\]
\end{lemma}

\begin{proof}
By the min-max principle,
\[
    \mu_n(T)
    =
    \inf_{\substack{L\subset\mathcal K\\ \dim L=n-1}}
    \sup_{\substack{\psi\in L^\perp\\ \|\psi\|=1}}
    ( \psi,T\psi) .
\]
For every $\psi\in\mathcal H$ with $\|\psi\|=1$, we have
$
    (\psi,T\psi)
    \leq
    (\psi,S\psi)+\|T-S\|$. 
Thus, for every subspace $L\subset\mathcal H$ with $\dim L=n-1$,
\[
    \sup_{\substack{\psi\in L^\perp\\ \|\psi\|=1}}
    (\psi,T\psi)
    \leq
    \sup_{\substack{\psi\in L^\perp\\ \|\psi\|=1}}
    (\psi,S\psi)
    +
    \|T-S\|.
\]
Taking the infimum over all such subspaces $L$ yields
$
    \mu_n(T)
    \leq
    \mu_n(S)+\|T-S\|$. 
Interchanging $T$ and $S$, we also obtain
$
    \mu_n(S)
    \leq
    \mu_n(T)+\|T-S\|$. 
Therefore the lemma follows. 
\end{proof}

\begin{corollary}\label{col:eigenvalues_conv}
Suppose that $|\alpha|\ne|\beta|$. 
Let $n\in\NN_0$ and $\#\in\{+,-\}$, and let $E_{n,\#}(g)$ denote the $n$-th eigenvalue of 
$K_{\alpha,\beta}(g)\upharpoonright_{\cH _\#}$, 
arranged in nondecreasing order and counted with multiplicity.
\begin{enumerate}
\item[(1)] If $|\alpha|>|\beta|$, then
\begin{align*}
    \lim_{|g|\to\infty}
    \left(
        E_{n,\#}(g) + \frac{g^2\alpha^2}{2}
    \right)
    =
    \sqrt{1 - \frac{\beta^2}{\alpha^2}} 
    \left(n + \frac12\right).
\end{align*}
\item[(2)] If $|\alpha|<|\beta|$, then
\begin{align*}
    \lim_{|g|\to\infty}
    \left(
        E_{n,\#}(g) + \frac{g^2\beta^2}{2}
    \right)
    =
    \sqrt{1 - \frac{\alpha^2}{\beta^2}}
    \left(n + \frac12\right).
\end{align*}
\end{enumerate}
\end{corollary}
\begin{proof}
We prove (1). The proof of (2) is similar. 
Let
$
    c = |\triangle| + \frac{|\beta|}{2|\alpha|}
$
as in Lemma~\ref{prop:uniform-lower-bound}. Note that $(\hat{K}_{\alpha,\beta,\#}(g) + c)^{-1}$ and $(\hat{h}_\infty + c)^{-1}$ are positive compact self-adjoint operators. By Theorem~\ref{thm:parity-reduction},
\begin{align*}
    \lim_{|g|\to\infty}
    \left\|
        (\hat{K}_{\alpha,\beta,\#}(g) + c)^{-1} - (\hat{h}_\infty + c)^{-1}
    \right\|
    = 0.
\end{align*}
Applying Lemma \ref{ev}, we obtain
\begin{align*}
    \lim_{|g|\to\infty}
    \mu_n \left( (\hat{K}_{\alpha,\beta,\#}(g) + c)^{-1} \right)
    =
    \mu_n \left( (\hat{h}_\infty + c)^{-1} \right).
\end{align*}
Using
\begin{align*}
    \mu_n \left( (\hat{K}_{\alpha,\beta,\#}(g) + c)^{-1} \right)
    =
    \frac{1}{
        E_{n,\#}(g) + g^2\alpha^2 / 2 + c
    }
\end{align*}
and
\begin{align*}
    \mu_n \left( (\hat{h}_\infty + c)^{-1} \right)
    =
    \frac{1}{\sqrt{1 - \beta^2 / \alpha^2} (n + 1/2) + c},
\end{align*}
we obtain the assertion of (1).
\end{proof}
The critical lines have a qualitatively different spectral limit.
\begin{corollary}\label{col:critical_spectrum}
Suppose that $|\alpha|=|\beta|\ne0$. For every $\lambda\in[0,\infty)$ and $\#\in\{+,-\}$, there exists  eigenvalues
\begin{align*}
    E_{\lambda,\#}(g)
    \in
    \operatorname{spec} 
    \left(
        K_{\alpha,\beta}(g) \upharpoonright_{\cH _\#}
    \right)
\end{align*}
for all sufficiently large $|g|$, such that
\begin{align*}
    \lim_{|g|\to\infty}
    \left(
        E_{\lambda,\#}(g) + \frac{g^2\alpha^2}{2}
    \right)
    = \lambda.
\end{align*}
\end{corollary}
\begin{proof}
By Theorem~\ref{thm:parity-reduction}, $\hat{K}_{\alpha,\beta,\#}(g)$ converges  to $q^2/2$ in the strong resolvent sense. 
Let $\lambda\in\operatorname{spec}(q^2/2)=[0,\infty)$ and define
\begin{align*}
    d_g 
    = 
    \operatorname{dist} 
    \left(
        \lambda, \operatorname{spec} (\hat{K}_{\alpha,\beta,\#}(g))
    \right).
\end{align*}
It is enough to show that $\lim_{|g|\to\infty}d_g=0$. Fix arbitrary $\eps>0$ and choose a unit vector
$
    \psi
    \in
    \operatorname{Ran} \one_{(\lambda - \eps, \lambda + \eps)}(q^2 / 2)$. 
By the spectral theorem,
\begin{align*}
    \operatorname{Im}
    \left(\psi, (\hat{K}_{\alpha,\beta,\#}(g) - \lambda - i)^{-1}\psi\right)
    =
    \left(\psi, ((\hat{K}_{\alpha,\beta,\#}(g) - \lambda)^2 + 1)^{-1}\psi\right)
    \le 
    \frac{1}{d_g^2 + 1},
\end{align*}
whereas
\begin{align*}
    \operatorname{Im}
    \left(\psi, \left(\frac{q^2}{2} - \lambda - i\right)^{-1}\psi\right)
    =
    \left(\psi, \left(\left(\frac{q^2}{2} - \lambda\right)^2 + 1\right)^{-1}\psi\right)
    \ge 
    \frac{1}{\eps^2 + 1}.
\end{align*}
Therefore, strong resolvent convergence implies that
\begin{align*}
    \frac{1}{\eps^2 + 1}
    \le
    \liminf_{|g|\to\infty} \frac{1}{d_g^2 + 1}
    =
    \frac{1}{(\limsup_{|g|\to\infty} d_g)^2 + 1}.
\end{align*}
Hence $\limsup_{|g|\to\infty} d_g \le \eps$. Since $\eps>0$ is arbitrary, we conclude that $\lim_{|g|\to\infty}d_g=0$.
\end{proof}
\subsection*{Acknowledgements}
The authors would like to thank Toshihide Hinokuma for his helpful simulation experiments, 
which were very informative for this paper.  This work is financially supported by JSPS KAKENHI Grant Numbers JP20K20886, JP23K20217, JP25H00595, and JP26K21807.
The authors also gratefully acknowledges the support of the Quantum and
Spacetime Research Institute (QuaSR), Kyushu University.%
\bibliographystyle{plain}
\bibliography{sample}
\end{document}